\documentstyle[epsfig,epsf,rotate]{mn}

\begin{document}

\title[XTE~J1859+226] 
{The X--ray transient XTE~J1859+226 in Outburst \&
Quiescence}
\author[C. Zurita et al.]
       {C. Zurita$^1$\thanks{e-mail: czurita@ll.iac.es}, 
	C. S\'anchez--Fern\'andez$^2$,
	J. Casares$^1$,
	P.A. Charles$^3$,
	T.M. Abbott$^4$,
\and
	P. Hakala$^5$,
	P. Rodr\'{\i}guez--Gil$^1$,
	S. Bernabei$^6$,
	A. Piccioni$^7$,
	A. Guarnieri$^7$,
	C. Bartolini$^7$,
\and
	N. Masetti$^8$,
	T. Shahbaz $^1$,
	A. Castro--Tirado$^9$\\
$^1$Instituto de Astrof\'\i{}sica de Canarias, 38200 La Laguna,
    Tenerife, Spain\\
$^2$Laboratorio de  Astrof\'\i{}sica Espacial y F\'\i{}sica Fundamental
(LAEFF-INTA), 28080  Madrid,  Spain\\
$^3$Department of Physics and Astronomy, University of Southampton, 
    Southampton, SO17 1BJ, UK\\
$^4$Nordic Optical Telescope, 38700 Santa Cruz de La Palma,
                Spain.\\
$^5$Observatory and Astrophysics Laboratory, FIN-00014 University of
       Helsinki, Finland \\
$^6$Osservatorio Astronomico di Bologna, Via Ranzani 1 Bologna, Italy\\
$^7$Dipartimento di Astronomia, Universit\'a di Bologna, Via Ranzani 1 Bologna, Italy\\
$^8$Istituto Tecnologie e Studio Radiazioni Extraterrestri, CNR, Via Gobetti 101, 40129 Bologna, Italy\\
$^9$Instituto de Astrof\'\i{}sica de Andaluc\'{\i}a, 18080 Granada, Spain \\
} 
\maketitle
%
%
\begin{abstract} 
We present  optical photometry and  spectroscopy of the X--ray  transient XTE
J1859+226,   obtained   during  outburst   and   its   subsequent  decay   to
quiescence. Both the X--ray and  optical properties are very similar to those
of  well--studied black  hole soft  X--ray  transients.  We  have detected  3
minioutbursts,  when XTE J1859+226  was approaching  quiescence, as  has been
previously  detected in  the  Soft  X--Ray Transients  GRO  J0422+32 and  GRS
1009--45.   By   24  Aug  2000   the  system  had  reached   quiescence  with
R=22.48$\pm$0.07.   The estimated  distance to  the source  is  $\sim$11 kpc.
Photometry taken during quiescence shows  a sinusoidal modulation with a peak
to peak amplitude  of about 0.4 mag. A period  analysis suggests that periods
from 0.28 to 0.47 days are equally possible at the 68\% confidence level. The
amplitude of the quiescent light curve and the relatively low ratio of X--ray
to optical flux,  indicates that the binary inclination  should be high.  The
measured colours during the outburst allows us to obtain the basic properties
of the disc, which agrees well with irradiated disc model predictions.

\end{abstract} 
\begin{keywords} black
hole physics--binaries:close--stars:  individual: XTE J1859+226--X-rays:stars
\end{keywords}
%
%
\section{Introduction} 
Soft X-ray  transients  (SXTs) are a  subclass   of low-mass  X-ray  binaries
(LMXBs) that are characterized  by episodic X-ray outbursts  (usually lasting
for several months), when the X-ray luminosities can increase by as much as a
factor of 10$^7$ \cite{vanparadijs95}. The observed optical flux is generated
by  X-ray  reprocessing in the accretion  disc  and the companion star. These
outbursts recur on a timescale of decades, but in the interim the SXTs are in
a state of quiescence and the optical  emission is dominated by the radiation
of the faint companion star. This offers  the best opportunity to analyze the
properties  of this star and  obtain  dynamical information which  eventually
enables us  to  constrain the   nature  of the   compact object.  There   are
currently 17 SXTs  with  identified optical counterparts,  with  13 dynamically
studied black-holes and 4 confirmed neutron stars.

The X--ray transient J1859+226, was  discovered by the All--Sky Monitor ({\it
ASM}) on board the Rossi X--Ray  Timing Explorer ({\it RXTE}) on 1999 October
9  \cite{wood99} and  subsequently  identified with  a  R=15.1 variable  star
\cite{garnavich00}.   Spectra during  outburst showed  typical  LMXB features
\cite{wagner99}. The  dereddened optical energy  distribution was represented
by a  steep blue power  law flattening in  the ultraviolet which  suggested a
binary  period $\le$1 day.   Although no  persistent coherent  modulation was
detected, a low amplitude (1 per cent modulation) with a period of 22--23 min
was  seen \cite{hynes99}.  Subsequent  photometry revealed  other modulations
with   periods    of   0.2806   days   \cite{uemura99}    and   0.3813   days
\cite{garnavich99}, but none  of them could be confirmed.   Nine months after
the outburst, but still not in quiescence, a new 0.78 day period was reported
\cite{mcclintock00}.  Light  curves folded  on this period  showed a  0.2 mag
primary minimum  (interpreted as a partial  eclipse of the  accretion disc by
the secondary star), and a shallow secondary minimum (explained as an eclipse
of the star by the disc).\\

Here we present our extensive  optical photometry of the entire outburst, the
subsequent minioutburst  and the first  quiescent photometry.  Some  of these
results have received  preliminary announcements in S\'anchez--Fern\'andez et
al. 2000, Zurita et al. 2000 and Charles et al. 2000. \\
%
%
\section{Observations and Data Reduction}
\begin{table}
\caption{Log of photometric observations (Oct 1999 --May 2000)
\label{log}}
\begin{tabular}{lcrc}
\hline 
\hline
{\em Date} & {\em HJD$^{(*)}$} & {\em Exp/Filter} & {\em Telescope}\\ 
\hline 
{\it Oct15}   &  7  & 1xR                      & 0.8m IAC80\\
{\it Oct17}   &  9  & 10xI,21xR,2xV,2xB        & 0.8m IAC80\\
{\it Oct18}   & 10  & 111xR,1xV,1xB            & 0.8m IAC80\\
{\it Oct19}   & 11  & 106xR,1xV,1xB            & 0.8m IAC80\\
{\it Oct20}   & 12  & 1xI,110xR,1xV ,1xB       & 0.8m IAC80\\
{\it Oct21}   & 13  & 2xR,1xV,1xB              & 0.8m IAC80\\
{\it Oct23}   & 15  & 1xR                      & 0.8m IAC80\\
{\it Oct28}   & 20  & 3xR                      & 0.8m IAC80\\
{\it Oct29}   & 21  & 1xI,111xR,1xV,1xB        & 0.8m IAC80\\
{\it Nov07}	 & 29  & 128xR	               & 1.5m OAN\\
{\it Nov27}   & 50  & 30xR,1xV,1xB	       & 0.8m IAC80\\
{\it Nov28}   & 51  & 40xR,1xV,1xB	       & 0.8m IAC80\\
{\it Dic10}   & 63  & 2xR                      & 0.8m IAC80\\
{\it Dic12}   & 65  & 4xR,1xV,1xB              & 0.8m IAC80\\
{\it Dic16}   & 69  & 1xR                      & 0.8m IAC80\\
{\it Jan03}   &  87  & 3xR              & 2.2m CAHA\\
{\it Jan04}   &  88  & 3xV              & 2.2m CAHA\\
{\it Jan05}   &  89  & 1xB              & 2.2m CAHA\\
{\it Feb01}   & 116  & 1xR              & 0.8m IAC80\\
{\it Feb04}   & 119  & 1xI,1xR,1xV,1xB           & 1m OGS\\
{\it Feb05}   & 120  & 1xI,1xR,1xV,1xB            & 1m OGS\\
{\it Feb06}   & 121  & 1xI,2xR,1xV,1xB            & 1m OGS\\
{\it Feb07}   & 122  & 1xI,1xR,1xV,1xB            & 1m OGS\\
{\it Feb08}   & 123  & 1xI,1xR,1xV,1xB            & 1m OGS\\
{\it Feb09}   & 124  & 1xI,1xR,1xV,1xB            & 1m OGS\\
{\it Feb10}   & 125  & 1xI,1xR,1xV,1xB            & 1m OGS\\
{\it Feb11}   & 126  & 2xI,2xR,2xV,2xB            & 1m OGS\\
{\it Feb12}   & 127  & 1xI,2xR,1xV,1xB          & 1m OGS\\
{\it Feb13}   & 128  & 2xI,2xR,1xV,1xB            & 1m OGS\\
{\it Feb14}   & 129  & 3xI,2xR,1xV,1xB            & 1m OGS\\
{\it Feb15}   & 130  & 2xI,1xR,2xV,2xB            & 1m OGS\\
{\it Feb20}   & 135  & 1xR              & 1m OGS\\
{\it Feb27}   & 142  & 1xR              & 0.8m IAC80\\
{\it Mar03}   & 145  & 1xR              & 0.8m IAC80\\
{\it Mar16}   & 160  & 1xV,1xR              & 0.8m IAC80\\
{\it Mar23}   & 167  & 1xV,1xR              & 0.8m IAC80\\
{\it Mar24}   & 167-8  & 10xR           & Lowell\\
           	&    & 1xV,1xR          & 0.8m IAC80\\
{\it Mar25}   & 168  & 1xR              & Lowell\\
{\it Mar27}   & 170-1  & 30xR           & Lowell\\
{\it Apr03}   & 178  & 1xV,1xR              & 0.8m IAC80\\
{\it Apr04}   & 179  & 1xV,1xR              & 0.8m IAC80\\
{\it Apr05}   & 180  & 1xR              & 0.8m IAC80\\
{\it Apr10}   & 185  & 1xV,1xR              & 0.8m IAC80\\
{\it Apr16}   & 191  & 1xR              & 0.8m IAC80\\
{\it Apr25}   & 200  & 1xV,1xR              & 0.8m IAC80\\
{\it Apr28}   & 203  & 2xR              & 0.8m IAC80\\
{\it May16}   & 221  & 2xR              & 0.8m IAC80\\
{\it May17}   & 222  & 1xR              & 0.8m IAC80\\
{\it May23}   & 228  & 2xR              & 1.52 Loiano\\
{\it May24}   & 229  & 2xR              & 1.52 Loiano\\
{\it May25}   & 230  & 2xR              & 0.8m IAC80\\
	      &      & 3xR      	& 1.52 Loiano\\
\hline 
\hline
\end{tabular}
\\

{\footnotesize $^*$HJD--2451460}\\
\end{table}
\normalsize
\begin{table}
\contcaption{Log of photometric observations (Jun--Nov 2000)}
\begin{tabular}{lcrc}
\hline 
\hline
{\em Date} & {\em HJD$^{(*)}$} & {\em Exp/Filter} & {\em Telescope}\\ 
\hline 
{\it Jun01}   & 237  & 37xR             & 1m OGS\\ 
{\it Jun03}   & 239  & 34xR             & 1m OGS\\ 
{\it Jun04}   & 240  & 64xR             & 1m OGS\\ 
{\it Jun06}   & 242  & 69xR             & 1m OGS\\ 
{\it Jun07}   & 243  & 28xR             & 1m OGS\\ 
{\it Jun11}   & 246  & 1xR              & 0.8m IAC80\\
{\it Jun12}   & 248  & 1xI,1xV          & 2.5 NOT\\
{\it Jun14}   & 249  & 1xV,1xR              & 0.8m IAC80\\
{\it Jun24}   & 260  & 1xR              & 0.8m IAC80\\
	      &      & 18xR             & 1m OGS\\ 
{\it Jun25}   & 261  & 17xR             & 1m OGS\\ 
{\it Jun28}   & 264  & 17xR             & 1m OGS\\ 
{\it Jul01}   & 267  & 1xV,1xR              & 0.8m IAC80\\
{\it Jul02}   & 268  & 2xR              & 1.52 Loiano\\
{\it Jul03}   & 269  & 1xR              & 1.52 Loiano\\
{\it Jul04}   & 270  & 1xV,1xR              & 0.8m IAC80\\
{\it Jul05}   & 271  & 336xB              & 2.5 NOT\\
{\it Jul06}   & 272  & 1xV,1xR              & 0.8m IAC80\\
	      &      & 887xB              & 2.5 NOT\\
{\it Jul11}   & 277  & 1xV,1xR              & 0.8m IAC80\\
{\it Jul14}   & 280  & 1xV,1xR              & 0.8m IAC80\\
{\it Jul15}   & 281  & 1xV,18xR              & 0.8m IAC80\\
{\it Jul16}   & 282  & 1xR             & 1m JKT\\ 
{\it Jul17}   & 283  & 1xR              & 0.8m IAC80\\
{\it Jul19}   & 285  & 1xR              & 0.8m IAC80\\
{\it Jul20}   & 286  & 1xR             & 1m JKT\\
{\it Jul21}   & 287  & 1xR              & 0.8m IAC80\\
	      &      & 1xR              & 1.52 Loiano\\
{\it Jul22}   & 288  & 1xR              & 0.8m IAC80\\
{\it Jul23}   & 289  & 1xR              & 1.52 Loiano\\
{\it Jul24}   & 290  & 1xR              & 1.52 Loiano\\
{\it Jul27}   & 293  & 1xR              & 1.52 Loiano\\
{\it Jul30}   & 296  & 168xR             & 1m OGS\\ 
{\it Jul31}   & 297  & 59xR             & 1m OGS\\
{\it Aug05}   & 302  & 1xR              & 0.8m IAC80\\
{\it Aug06}   & 303  & 1xR              & 0.8m IAC80\\
	      &      & 1xR              & 1.52 Loiano\\
{\it Aug07}   & 304  & 1xR              & 1.52 Loiano\\
{\it Aug08}   & 305  & 1xV,1xR              & 0.8m IAC80\\
{\it Aug09}   & 306  & 1xV,1xR              & 0.8m IAC80\\
{\it Aug10}   & 307  & 1xR              & 1.52 Loiano\\
{\it Aug11}   & 308  & 1xR              & 1.52 Loiano\\
{\it Aug18}   & 315  & 1xR              & 1.52 Loiano\\
{\it Aug19}   & 316  & 1xR              & 1.52 Loiano\\
{\it Aug20}   & 317  & 1xR              & 1.52 Loiano\\
{\it Aug21}   & 318  & 1xR              & 1.52 Loiano\\
{\it Aug24}   & 321  & 1xR              & 0.8m IAC80\\
{\it Aug25}   & 322  & 1xR              & 0.8m IAC80\\
{\it Aug27}   & 324  & 1xR              & 0.8m IAC80\\
{\it Aug28}   & 325  & 1xR              & 0.8m IAC80\\
{\it Aug29}   & 326  & 1xR              & 0.8m IAC80\\
{\it Aug04}   & 332  & 1xR              & 0.8m IAC80\\
{\it Aug05}   & 333  & 1xR              & 0.8m IAC80\\
{\it Sep27}   & 355  & 6xR              & 4.2m WHT\\
{\it Sep28}   & 356  & 16xR             & 4.2m WHT\\
{\it Oct17}   & 374  & 1xV             & 2.5m NOT\\
{\it Nov04}   & 393  & 6xR             & 2.5m INT\\
{\it Nov05}   & 394  & 5xR             & 2.5m INT\\
\hline
\hline
\end{tabular}
\\
{\footnotesize $^*$HJD--2451460}\\
\end{table}

\subsection{Photometry}
Our  long--term  monitoring  campaign  was  carried  out  during  the  period
Oct. 1999--  Sep.  2000 with  the 80 cm  IAC80 and 1m Optical  Ground Station
(OGS) at the Observatorio del Teide; 1m  JKT, 2.5m NOT, 3.5m TNG and 4.5m WHT
at the  Observatorio del Roque  de los Muchachos;  1.5m OAN and 2.2m  CAHA at
Observatorio de  Calar Alto and 1.52m  telescope at Loiano.   We obtained CCD
images mainly in $R$--band but  also some $B,V,I$ colours.  Integration times
ranged  from  30~s  to  40~min  , depending  on  the  telescope,  atmospheric
conditions and the  star brightness.  The observing log  with full details is
given in  Table~\ref{log}.  All the images were  de--biased and flat--fielded
in the standard way using {\sc iraf}.

\begin{figure}
\begin{center}
\epsfig{file=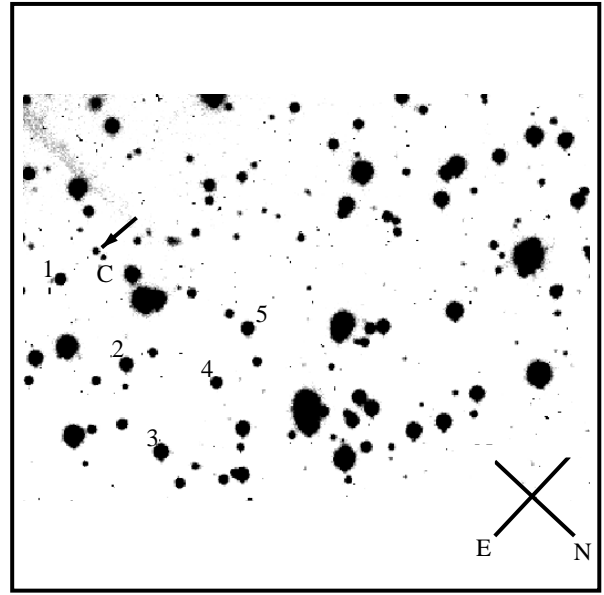,width=8cm,height=8cm,angle=0}
\caption{$R$ band  1800~s image of XTE J1859+226. The field
of view is 1.4  x 1.0 arcmin. XTE J1859+226 is indicated  by an arrow and the
contaminanting star by the letter C.  The magnitudes of stars 1--5 are listed
in Table~\ref{calib}
\label{field}}
\end{center}
\end{figure}

We applied  aperture photometry to  our object and several  nearby comparison
stars  within the  field  of view,  using  {\sc iraf}.   We selected  several
comparison stars  which were checked  for variability during each  night, and
over the  entire data  set. Calibration  of the data  was performed  using 17
standard stars  from several fields  \cite{landolt92}, to construct  a colour
dependent calibration.  Frames taken during  quiescence with the WHT and good
seeing conditions,  revealed the  presence of a  faint nearby  star $\sim$1.4
arsec  North  of  the   target  (see  Figure~\ref{field}).   We  applied  psf
photometry using  the {\sc IRAF} routine {\sc  DAOPHOT} \cite{stetson87}.  We
found a magnitude of $R$=23.05$\pm$0.04  for the contaminating star.  We also
recalibrated a set  of 5 faint comparison stars  from the previous calibrated
set, using these images (see Table~\ref{calib}).

\begin{table}
\caption{Log of spectroscopic observations
\label{log_spec}}
\begin{tabular}{lcrc}
\hline 
\hline
{\em Date} & {\em HJD$^{(*)}$} & {\em Number of spectra} & {\em Telescope}\\ 
\hline
{\it 99 Oct28}  & 20  &  1  &  1.52m G.D. Cassini     \\
{\it 99 Oct28}  & 20  &  7  &  2.5m  INT                   \\
{\it 99 Oct29}  & 21  &  24 &  2.5m  INT                   \\
{\it 00 Jul11}  & 277 &  12 &  4.5m  WHT  \\
\hline 
\hline
\end{tabular}\\
{\footnotesize $^*$HJD--2451460}\\
\end{table}
\normalsize

\subsection{Spectroscopy}
Spectroscopic  observations of  XTE  J1859+226 were  carried  out at  primary
outburst and  during one of the  minioutbursts.  The observing  log for these
observations  is  presented in  Table~\ref{log_spec}.   The  first series  of
optical spectra  of the source  were obtained on  Oct.  28-29, 1999,  when 31
exposures were taken in the  Isaac Newton Telescope (INT) at the Observatorio
del Roque de los Muchachos, using  the {\sc eev10} camera on the Intermediate
Dispersion Spectrograph ({\sc ids}) (range 3500-5000 \AA, spectral resolution
1.02 \AA/pixel).   The exposure times ranged from  300--600\,s for individual
spectra.  A single 2400 s spectrum  was obtained with the 152cm G.D.  Cassini
Telescope at Loiano  Observatory using {\sc bfosc}.  The  slit width was 2.5"
and the  spectral range  3500-9000 \AA.  The  second series  of spectroscopic
observations  were carried out  on July  11 2001,  during the  second optical
minioutburst   described  in   section~\ref{fotometria}.   We   observed  XTE
J1859+226 using the  {\sc isis} spectrograph on the  4.2m WHT in Observatorio
del Roque de los Muchachos, using typical exposure times of 1800 s.  Standard
{\sc iraf}  procedures were  used to  de--bias the images  and to  remove the
small  scale CCD sensitivity  variations.  One  dimensional spectra  were then
extracted  from the  processed  images using  the  optimal extraction  method
(Horne    1986).    Wavelength    calibration   was    interpolated   between
contemporaneous exposures  of a copper--argon arc lamp.   No flux calibration
was performed on the spectra, instead continuum normalization was applied.

\begin{table}
\caption{$R$ magnitudes for the contaminating star and 5 faint comparison
in the field of J1859+226
\label{calib}}
\begin{tabular}{cccccc}
\hline 
\hline  
{{\it C}} & {{\it  1}} & {{\it 2}}  & {{\it 3}} &  {{\it 4}} &
{{\it  5}}\\  
\hline 
{23.05}  &  {20.31}  & {19.41}  &  {18.86}  & {20.15}  &
{19.58}\\ {$\pm$0.04} & {$\pm$0.06} & {$\pm$0.05} & {$\pm$0.05} & {$\pm$0.06}
& {$\pm$0.05}\\ 
\hline 
\hline
\end{tabular}
\end{table}

%
%
\section{Long term behaviour} 

\begin{figure*} 
\begin{center}
\epsfig{file=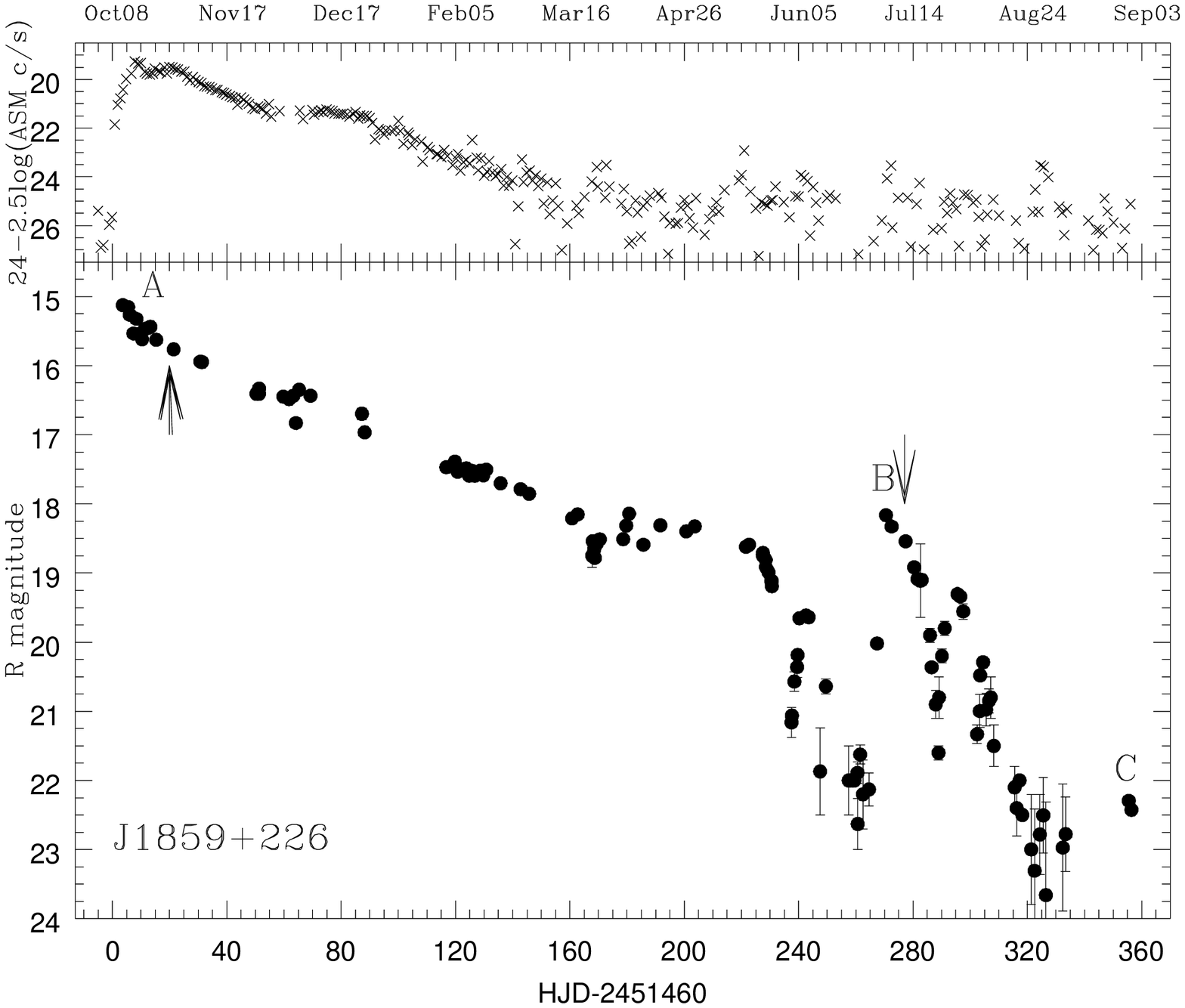,width=14cm,angle=0}
\epsfig{file=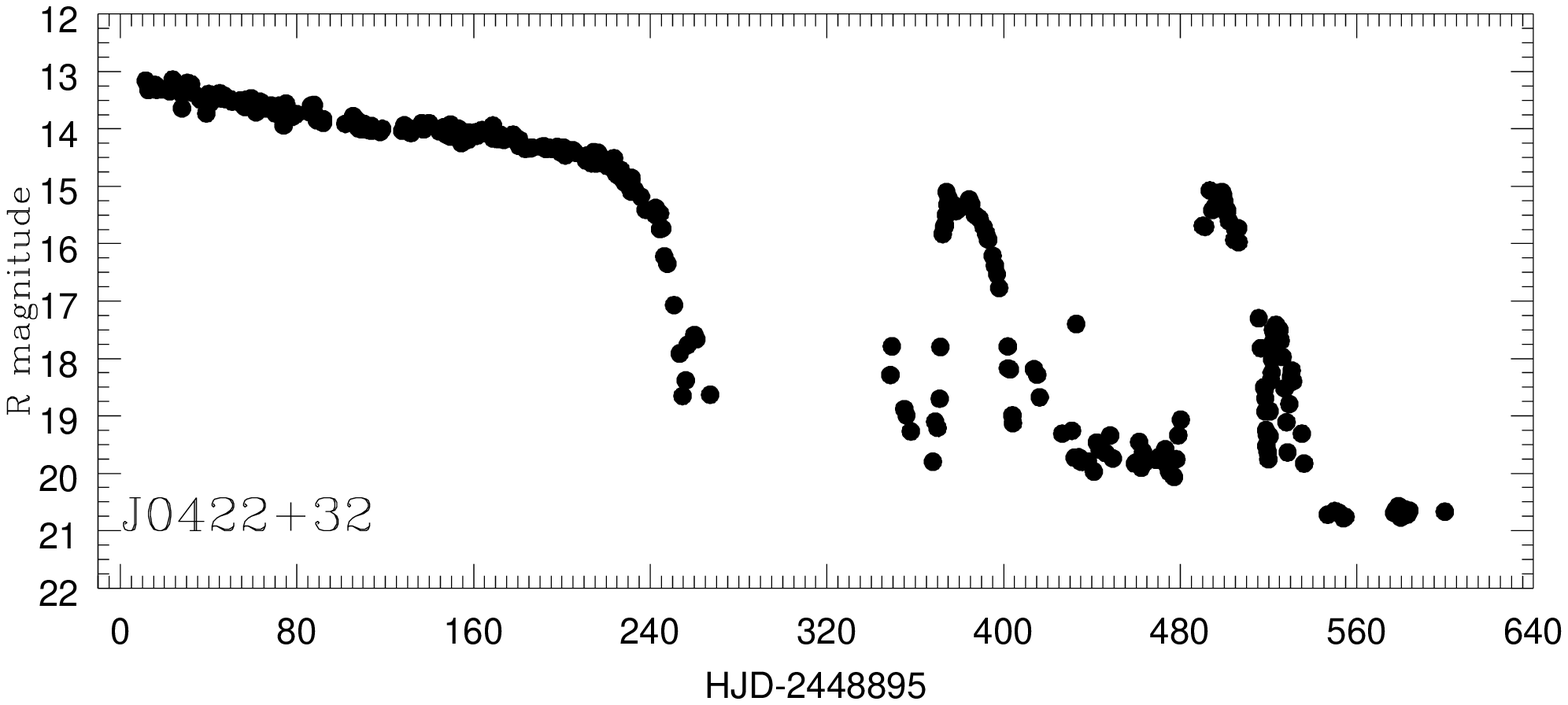,width=14cm,angle=0}
\caption{{\bf  Top}:  Temporal  evolution  of  J1859+226  plotted  as  'X-ray
magnitudes'  [24-2.5$log$(XTE/ASM   Count  Rate)]  and   $R$-band  magnitudes
averaged for each  day.  Note that 1 Crab equals an  {\sc xte/asm} count rate
of 75  counts/sec. A,  B and C  mark the  epochs studied for  photometric and
spectrometric variability .  The arrows  indicate the dates when spectra were
taken.   {\bf Bottom}:  $R$  band light  curve  of XTE  J0422+32 plotted  for
comparison.  The J0422+32 data were provided by E. Kuulkers.
\label{long_curve}}
\end{center}
\end{figure*}

\begin{figure*} 
\begin{center}
\epsfig{file=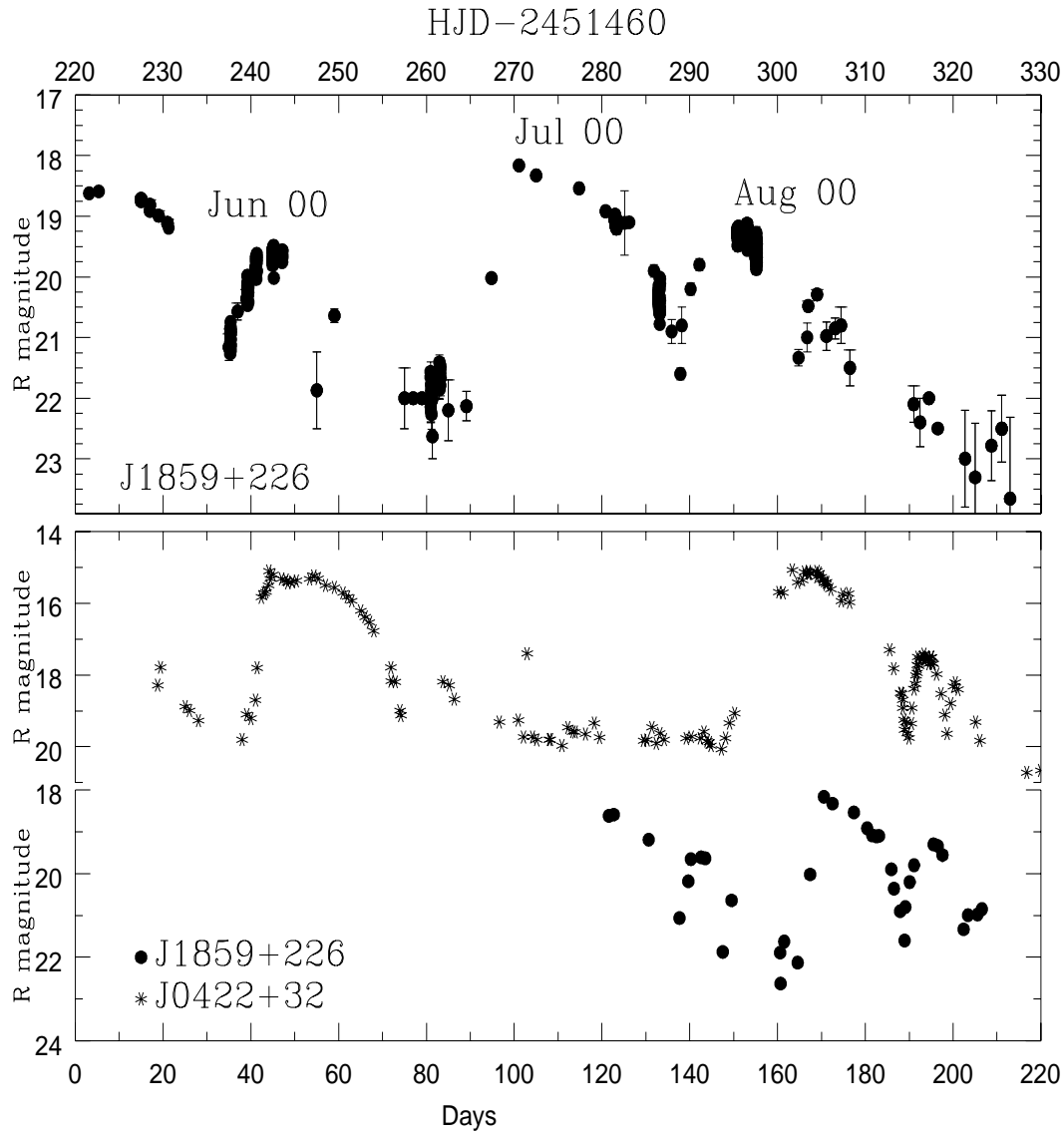,width=14cm,height=15cm,angle=0}
\caption{{\bf Top:} The decline of the main outburst of J1859+226 showing the
subsequent minioutbursts.  The three minioutbursts  are marked as Jun 00, Jul
00  and  Aug 00.   {\bf  Bottom:} The  minoutburst  of  J0422+32 plotted  for
comparison (asterisks).  We have  superimposed the J1859+226 minioutbursts on
the same scale  (filled circles). For both systems, the  last point marks the
day  when  quiescence  is  reached.   The  J0422+32  data  were  provided  by
E. Kuulkers.
\label{miniout}}
\end{center}
\end{figure*}

Both the X--ray and optical properties  of XTE J1859+226 were very similar to
those of  previously well--studied black hole  SXTs, and so we  embarked on a
campaign of systematic monitoring of the decay light curve.  Optical outburst
light curves of SXTs tend to be fragmentary and generally cover only the main
outburst.   We note  that  the only  SXT  that has  been extensively  covered
throughout the entire outburst and  subsequent decay activity is GRO J0422+32
(Callanan     et    al.     1995,     Chevalier    \&     Ilovaisky    1995).
Figure~\ref{long_curve} therefore  compares the overall 1--yr  light curve of
GRO J0422+32 with of XTE  J1859+226 in optical ($R$--band) and X--rays (2--12
keV), since the  X--ray turn--on of 9 October 1999.   The basic properties of
the outburst  light--curve, are summarized in  Table~\ref{resumen}.  They are
similar to the light curves reported  in Chen et al. 1997.  The optical light
curve  can  be classified  as  {\it  FRED}: 'fast--rise  exponential--decay',
characterized by a  smooth decline of 0.017 mag/day.  At  about the time that
the X--rays dropped below the RXTE sensitivity limit, the optical light curve
began a precipitous  fall from $R$$\sim$18.5 to $R$$\sim$22.5  (day 210 after
outburst). About 12 days after the peak of the outburst, the X--ray intensity
reaches a secondary maximum which is not visible in  our optical data.

A 'glitch' of $\Delta$$R$$\sim$0.5 mag  was detected in the exponential decay
phase, at  about Mar 25  ($\sim$165 days after  the peak), and in  the X--ray
light curve about 5 days  earlier ($\sim$160 days from the outburst). Similar
glitches  have also  been observed  in the  optical light  curve  of A0620-00
\cite{tsunemi77}, with a magnitude increase of $\Delta$$B$$\sim$0.5 mag about
150  days  after  the onset  of  the  outburst,  and  also in  J0422+32  with
$\Delta$$V$$\sim$0.3 mag \cite{callanan95}.  In X--rays, glitches are seen in
the   light   curves   of   J0422+32,  A0620-00,   1543-47,   and   1124--683
\cite{chen97}.

\begin{table}
\caption{Basic properties of the outburst.
\label{resumen}}
\begin{tabular}{lr}
\hline
\hline
{{\it Date of the outburst:}} & {Oct.1999}\\
\hline
{\em X--ray light curve}\\
\hline 
{{\it Morphology type:}} & {FRED}\\
{{\it Instrument and E band:}} & {ASM(RXTE)/2--12 KeV}\\
{{\it Flux at the peak:}} & {0.25 Crab}\\
{{\it Rise timescale ($\tau_r$):}} & {2.5 days}\\
{{\it Decay timescale ($\tau_d$):}} & {34 days}\\
{{\it Duration of the rising phase (T$_r$):}} & {13 days}\\
{{\it Duration of the decay phase (T$_d$):}} & {155 days}\\
\hline
{\em Optical light curve}\\
\hline
{{\it Morphology type:}} & {Exponencial decay}\\
                         & {No rise phase data}\\
{{\it Peak magnitude}} & {R=15.1,V=15.3 mags}\\
{{\it Quiescent magitude}} & {R=22.48, V=23.29 mags}\\
{{\it Outburst amplitude (m$_{quiet}$-m$_{peak}$)}} & {$\Delta$R=8.4, $\Delta$V=7.8 mags}\\
{{\it Decay timescale ($\tau_d$):}} & {103 days}\\
{{\it Duration of the decay phase (T$_d$):}} & {220 days}\\

\hline 
\hline
\end{tabular}
\end{table}

Shortly after this  rapid decline began, we have  detected 3 minioutbursts or
small  amplitude  (relative to  the  primary  peak)  events superposed  on  a
'normal' decay  profile, when XTE J1859--226 was  approaching quiescence (see
Figure.~\ref{miniout}).  The  first one (Jun 00) occurs  $\sim$240 days after
the peak, has an amplitude of 1.3  mag in $R$ (with respect to the base level
of Jun 2000) and lasts for  $\sim$20 days.  After the minioutburst, the decay
follows roughly the same decline as before.  The second minioutburst (Jul 00)
occurs  $\sim$265 days  after  the  peak, reaches  a  maximum of  $R$$\sim$18
($\Delta$R$\sim$4.7mag) and lasts  for $\sim$25 days. And the  third (Aug 00)
reaches a peak  only $\sim$2 mags above the  previous minimum.The source then
decays  linearly to  $R\sim$23 about  30 days  after the  onset on  this last
event.   Minioutbursts have  only been  previously detected  in  GRO J0422+32
\cite{chevalier95}  and  GRS  1009--45  \cite{bailyn92}  but this  may  be  a
consequence  of  inadequate monitoring  (or  lack  of  sensitivity) once  the
outburst  is over.  In  J0422+32, two  principal minioutbursts  were observed
(see Figure.\ref{long_curve}).   Both had an  amplitude of $\Delta$$V$$\sim$5
mag and lasted for $\sim$20--40 days. Here the minioutbursts reached the same
level as  the extrapolated  light curve before  the precipitous fall  and are
followed  by  events of  smaller  amplitude, as  occur  in  J1859+226 Jul  00
minioutburst.  Despite these  similarities, the shape of the  long term light
curve in  J0422+32 is completely  different.  Here the  rate of decay  in the
$R$--band  is  only   0.0056  mag/day,  which  is  a   remarkably  slow  rate
\cite{callanan95}. We  note that both the  duration of the  main outburst and
the minioutbursts is approximately the same.

\begin{figure} 
\begin{center}
\epsfig{file=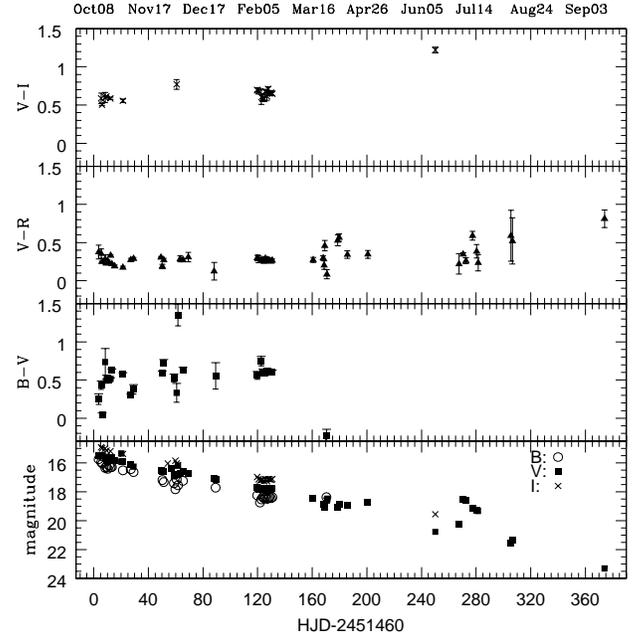,width=10cm,height=10cm,angle=0}
\caption{Long term variations of the $V--I$, $V--R$ and $B--V$ colour indices
of J1859+226 are plotted here against the overall light curve (bottom).
\label{color}}
\end{center}
\end{figure}

Figure~\ref{color} presents colour information  of J1859+226 as a function of
time.  The mean $V$--$R$ colour during outburst was measured to be $\sim$0.27
and  remains constant during  the main  outburst.  The  source does  not show
detectable  colour changes  during any  given night.   It is  clear  that the
system  reddens  as the  system  approaches  quiescence  and the  secondary's
contribution increases.   During quiescence $V$--$R$=0.81$\pm$0.11,  which is
consistent with the colour of a G9--K5 main sequence star.  \\

From  24 Aug  the system  has  reached quiescence  with a  mean magnitude  of
$R$=22.48$\pm$0.07 and $V$=23.29$\pm$0.09 (measured  on 17 Oct).  This yields
a total amplitude for the optical outburst of $\Delta$$V$=7.8 mag, comparable
to that observed for other SXTs \cite{chen97}.

%
%
\section{Analysis of photometric variability}
\label{fotometria}

\begin{figure} 
\begin{center}
\epsfig{file=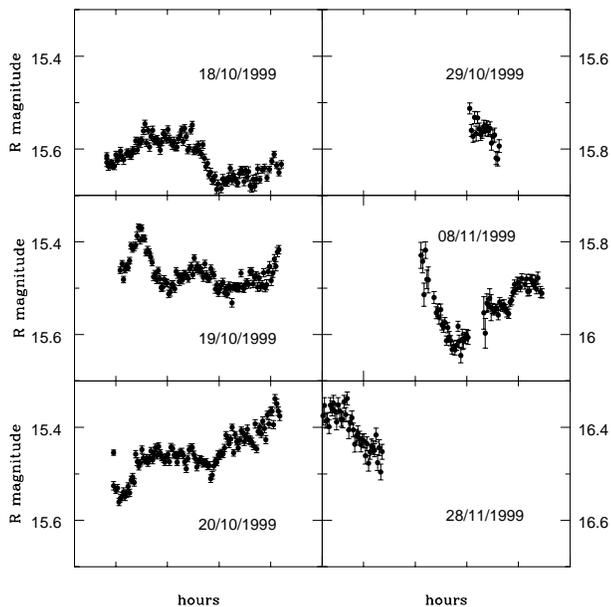,width=9cm,height=9cm,angle=0}
\caption{Optical light curves for XTE J1859+226 on different nights in Epoch
A. Note
the remarkable changes in the light curve shape during each night.
\label{short_curve}}
\end{center}
\end{figure}

Photometric variability has been explored in our best  sampled data set. This
includes a  group of 6 nights at   the peak of  outburst (18--20,  29 October
1999, 8   and 28 November   1999), 2 nights during the   maximum of  the July
minioutburst (5--6 July 2000) and   4 nights in quiescence (27--28  September
2000 and  4--5 November 2000). Hereafter, we  will refer to data taken during
the main outburst, the Jul 00 minioutburst and quiescence as  Epochs A, B and
C respectively and they are marked in Figure.~\ref{long_curve}.\\
 
\subsection{Epoch A: Irregular Variability}

Representative light  curves of   XTE  J1859+226, at   the beginning  of  the
outburst, are presented  in Figure~\ref{short_curve}. We note the  remarkable
changes   in  shape and   amplitude  with  no   evidence  for  repeatability.
Intermittent modulations of  this   type  during decline  from   outburst  on
timescales at   or  near the orbital  period   have been seen in  GS  2000+25
\cite{charles91},  Nova  Muscae 1991   \cite{bailyn92}  and Nova  Persei 1992
(Callanan et al.  1995; Chevalier and Ilovaisky 1994).\\

Fourier analysis of these October  data, shows three main peaks of comparable
significance   at   0.1600,   0.1907   and   0.2353   days   in   the   power
spectrum. However, no clear modulation  is obtained after folding the data on
these periods. We also note that  none of these periods have been observed by
other authors.  For example periods of 0.2806 days \cite{uemura99} and 0.3813
days \cite{garnavich99} were reported during the first phase of the outburst.
Nonetheless,  the determination  of orbital  periods of  X--ray  novae during
outburst has proved notoriously  unreliable, e.g.  modulation in the outburst
light curve at a period much shorter than the orbital period has been seen in
V404 Cyg (i.e. Wagner  et al. 1990, Gotthelf et al. 1991).  On the other hand
it has been suggested that J1859+226, has a binary period shorter than 1 day,
since the dereddened optical spectral energy distribution of J1859+226 during
outburst  can be  represented by  a steep  blue power  law flattening  in the
ultraviolet, and  this resembles energy distributions  of other short--period
SXTs \cite{hynes99} .

\subsection{Epoch B: 22 min QPOs}
\begin{figure*} 
\begin{center}
\epsfig{file=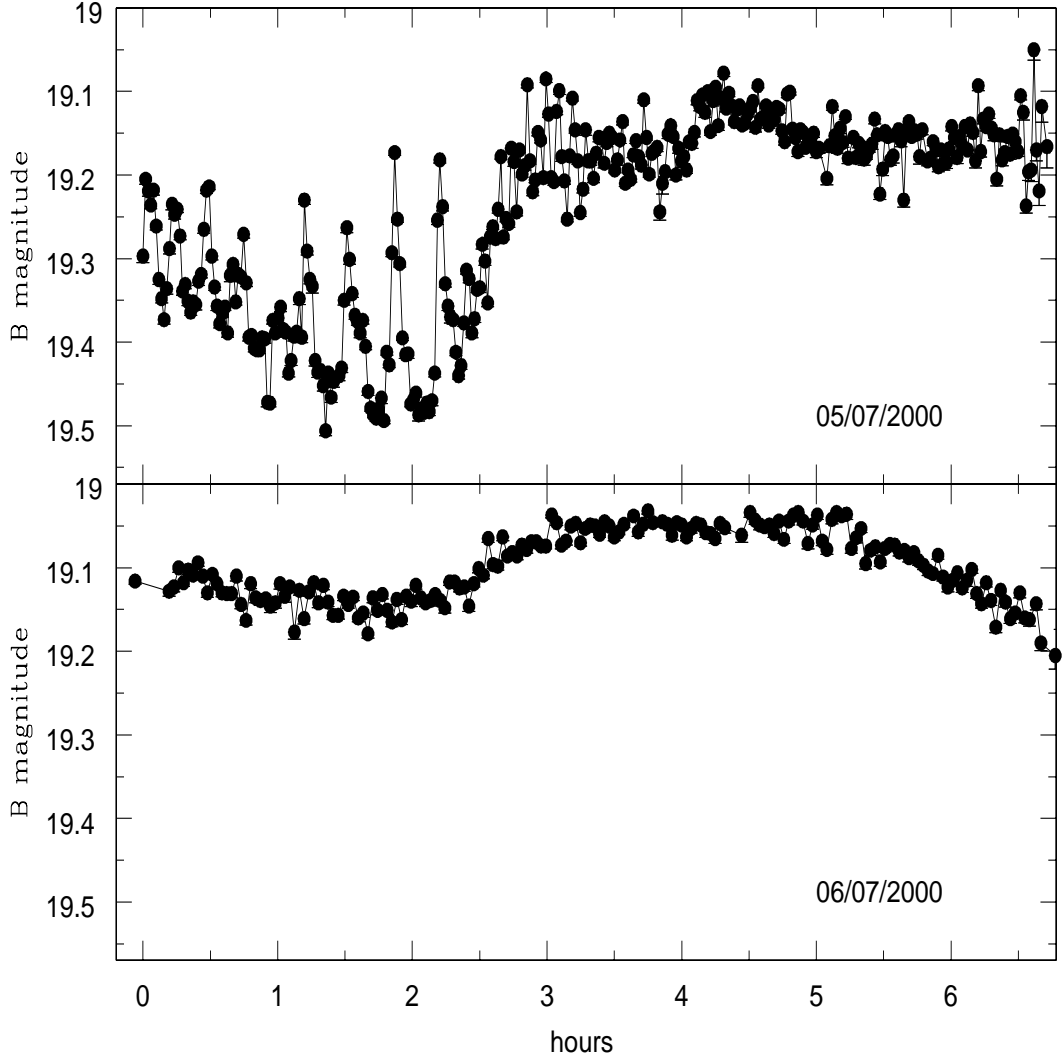,width=14cm,height=14cm,angle=0}
\caption{B--band light during  Epoch B (Jul00 minioutburs) of  J1859+226 on 5
({\bf top}) and 6 Jul ({\bf bottom}). The data points have been obtained with
the 2.5m NOT. Note the flares in the Jul 5 light curve.
\label{curv0506jul}}
\end{center}
\end{figure*}

High time resolution  (30 to 60 s) $B$--band light curves  were obtained on 5
and  6  July  when J1859+226  was  at  the  maximum  of a  minioutburst  (see
Figure.~\ref{curv0506jul}). The light curve on  Jul 5 shows an extended 'dip'
of about  0.3 mag lasting  for 3 hr.   During this dip, the  source exhibited
remarkable {\it QPO}  flaring activity with $\sim$20 per  cent amplitude.  On
July 6,  the {\it QPOs}  has disappeared and  instead the source  exhibited a
smooth  sinusoidal modulation  with $\sim$0.10  mag amplitude.   Studying the
rapid variability seen on July 5,  we find a period of 21.7$\pm$0.6 min using
a  periodogram analysis.  This  {\it QPO}  timescale  is very  close to  that
reported by Hynes et al.  1999.   The flaring ceased when the source returned
to its predip level.

A  periodicity of 0.78  days has  also been  reported from  contemporary data
\cite{mcclintock00}, which is claimed to  be the orbital period.  Their light
curve shows evidence  for a deep primary minimum  (interpreted as the eclipse
of the disc by the companion  star) and a shallow secondary minimum. Assuming
that the 0.3  mag dip in the July  05 light curve is the  primary minimum, it
occurs at HJD  2451731.51, which is exactly 1.3  days before McClintocks's et
al.   primary  minimum.  Therefore  the   true  orbital  period  would  be  a
submultiple  of 1.3 days,  which automatically  rules out  the 0.78  d period
\cite{zurita00}.

\subsection{Epoch C: Quiescent ellipsoidal modulation}

\begin{figure} 
\begin{center}
\epsfig{file=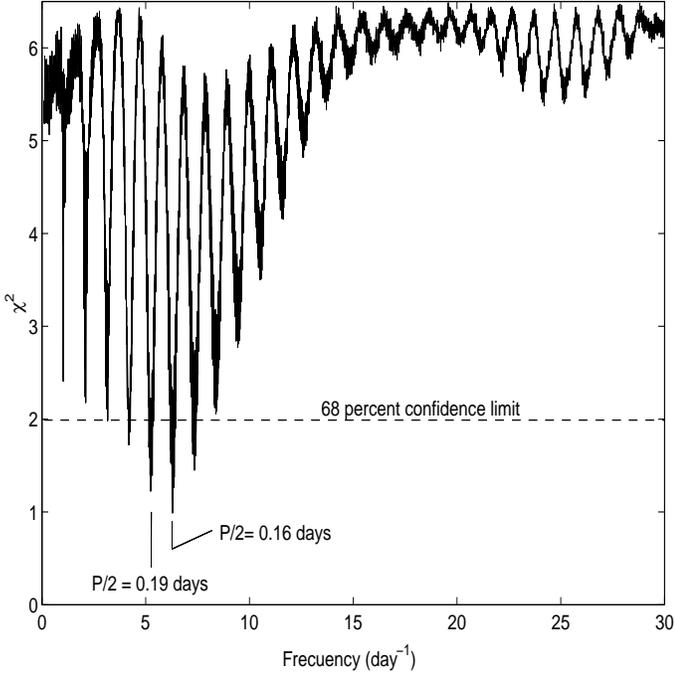,width=9cm,height=9cm,angle=0}
\caption{$\chi^2$ minimization  spectrum of  the quiescent data.  The deepest
minima are  marked, corresponding to  P$_{orb}\sim$0.32 and P$_{orb}\sim$0.38
days.
\label{chi2}}
\end{center}
\end{figure}

Data taken in quiescence during 2000  September 27 and 28, on the 4.2\,m WHT,
and 2000 November 4 and 5 on the 2.5\,m INT, exhibit a sinusoidal modulation,
with a  peak-to-peak amplitude  of about 0.4  mag.  We applied  a Chi-squared
minimization method in  order to characterize the periodicity  present in the
data.   Periodogram is  shown  in Figure~\ref{chi2}.   The periodogram  shows
several peaks  although formally  the peak with  the highest  significance is
found  at   0.1594$\pm$0.003  days.   Assuming   that  this  is   actually  a
double--humped  ellipsoidal light  curve,  as is  typical  of quiescent  soft
X--ray  transients, the orbital  period would  then be  0.3188$\pm$0.003 days
($\sim$7.7 hours).  We note that this  period is roughly a submultiple of 1.3
days.  This  suggests that  the variability (and  in particular the  0.35 mag
drop  observed in the  July minioutburst),  might be  related to  the orbital
period.  However, periods from 0.14 to 0.23 days (P$_{orb}\sim$6.6 to 11.2 hrs
) are equally possible at the  68 percent confidence level.  Clearly more and
higher quality data are needed  to determine conclusively which period is the
true orbital period.\\

We  note  that  Filippenko  \&  Chornock (2001)  report  a  9.16$\pm$0.08  hr
(0.382$\pm$0.03 days) modulation in the radial velocities of only 10 spectra,
spread over two nights, obtained during quiescence. This period is consistent
with  our  data  and  previous   reports  by  Garnavich  et  al.  (2001)  and
S\'anchez--Fern\'andez et al.  (2000). Since  all the dataset are affected by
aliasing, it  is clear  that more spectroscopic/photometric  observations are
required to determine the true orbital period.\\

\begin{figure} 
\begin{center}
\epsfig{file=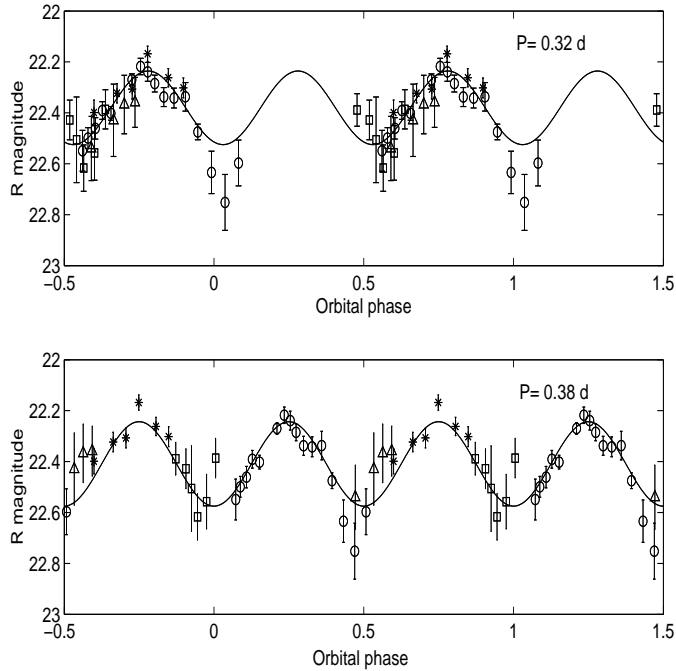,width=9cm,height=9cm,angle=0}
\caption{Optical  photometry  of XTE  J1859+226  during  the quiescent  state
(Epoch  C) folded  on the  0.319\,d period  (top panel)  and on  the 0.382\,d
period  found by  Filippenko \&  Chornock (bottom  panel) and  a superimposed
sinusoidal fits.   We have represented  points taken on  different days/sites
with different symbols: asterisks  (27/09/2000 -- WHT), circles (28/09/2000--
WHT), squares (04/11/2000 -- INT) and triangles (05/11/2000 -- INT).
\label{elip}}
\end{center}
\end{figure}

    In Figure~\ref{elip} we show the  quiescent data folded on the 0.319 days
period (top panel) and  on the 0.382$\pm$0.03 days Filippenko's spectroscopic
period (bottom  panel).  By comparing the  amplitude of the  curve with other
SXTs in quiescence, we conclude that the the binary inclination must be high.
The different depths in the minima  also support this idea since this is only
evident when the inclination angle is high.

An  alternative  way  to estimate  the  orbital  period  is through  the  $V$
amplitude  of the  outburst  light curve  ($\Delta$$V$),  by the  use of  the
empirical relation
$$    \Delta    V=14.36-7.6\log{P_{orb}}(hr)$$   \cite{shahbaz98}.     Taking
$V$(peak)=15.5 and $V$(quiescence)=23.3, we obtain P$_{orb}\sim$7.30 hr (0.30
days)  which is  consistent with the previous values.

%
%
\section{Analysis of spectral variability}
\begin{figure*} 
\begin{center}
\epsfig{file=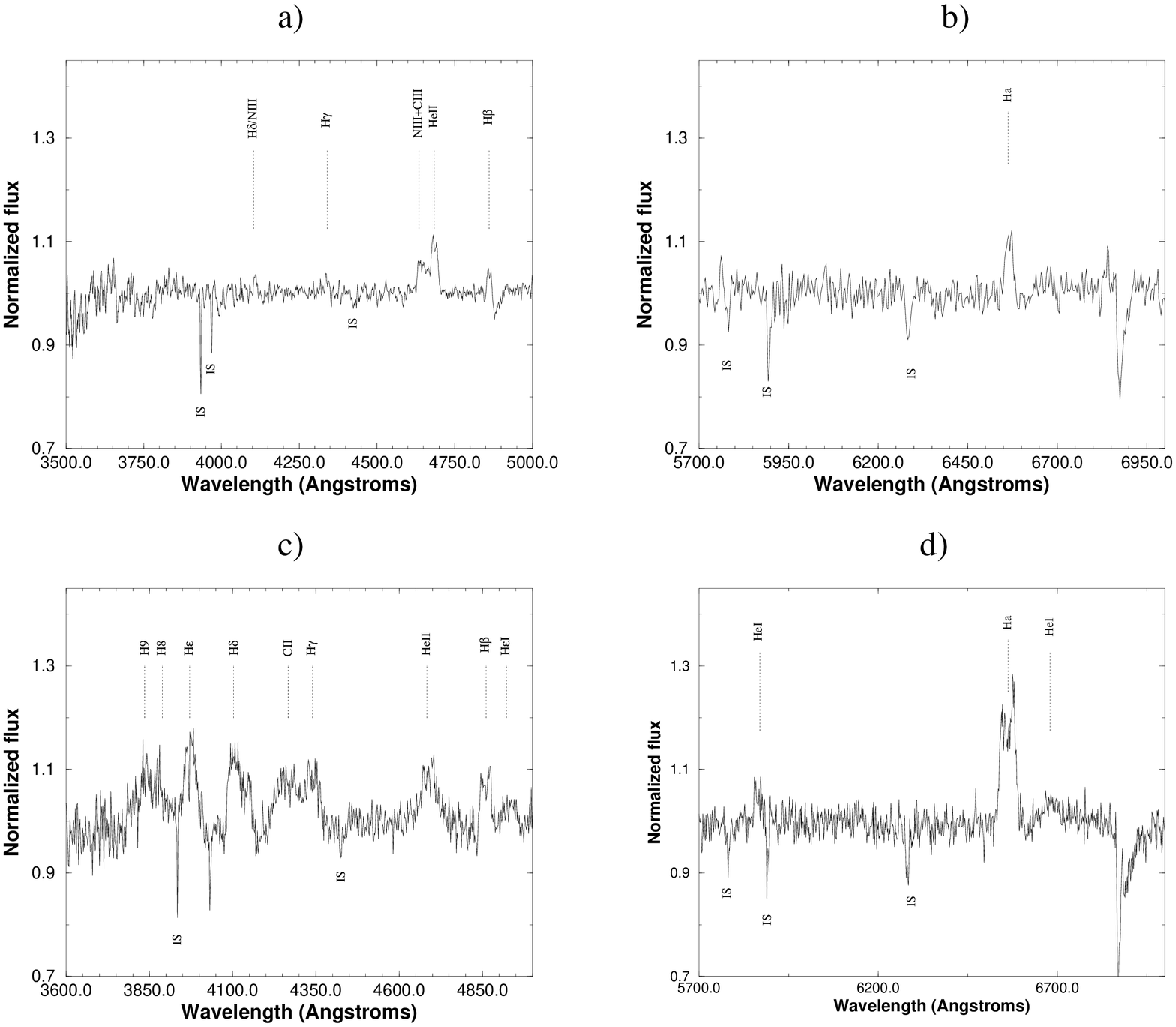,width=15cm,height=15cm,angle=0}
\caption{Spectroscopic observations  of XTE J1859+226 during  Epoch A (main
outburst)--   figures   a  and  b--    and    during   Epoch  B   (July    00
minioutburst)--figures c and d.
\label{spec}}
\end{center}
\end{figure*}

Spectral variability has  been  explored in epoch A  (Oct.20--28)  and in the
epoch B minioutburst  (Jul.11).  Days when  spectra were taken are marked  by
arrows in Figure.~\ref{long_curve} and are plotted in Figure.~\ref{spec}.

\subsection{Epoch A}

The bluer spectrum  displayed typical emission lines of  X--ray transients in
outburst: He {\sc ii}  4686 \AA, the Bowen blend at  4630--4640 \AA and 3097--
4103 \AA, and Balmer emission  from H${\beta}$ to H${\delta}$. The H${\beta}$
emission line is embedded in a  broad absorption, as is often observed in the
Balmer  emission lines  from SXTs  (e.g.  Casares  et al.   1995).   The main
interstellar features were  the Ca{\sc ii} interstellar bands  at 3933.66 and
3968.47 \AA,  (probably with some contribution  from H${\epsilon}$ absorption
at 3970 \AA), and the diffuse interstellar band at 4430 \AA.
 
The only  prominent feature in  the redder spectrum is  H${\alpha}$ emission.
The main interstellar features were the blend at 5778, 5780 and 5797 \AA, the
blend due  to the interstellar  Na I D  lines at 5889  and 5895 \AA,  and the
interstellar band at 6280 \AA.  The HeII and Balmer lines appear to be double
peaked with separations  of 300--500 km s$^{-1}$, notably  smaller than those
detected in the  outburst spectra of GRO J0432+22 (Callanan  et al.  1995) or
in  the quiescent  spectra of  Nova Muscae  1991 or  A0620-00 \cite{orosz94}.
Double peaks were not detected in  the Bowen blend lines in our spectra, most
probably  due  to  the  blending  phenomena  than  to  the  absence  of  such
structure.  However, the  Bowen blend  could also  arise from  other emitting
region in the system.

\subsection{Epoch B}
The second  series of spectra  of the source  were obtained on July  11 2000,
near the maximum of that  minioutburst. The bluer spectra displayed prominent
Balmer  emission from  H${\beta}$  to H$9$.   Bowen  emission was  not
detected in  these observations.   On the other  hand, C{\sc ii}  emission at
4267  \AA  was  present.   The  main interstellar  features  were  again  the
interstellar absorption at 3968 \AA, which affects the H${\epsilon}$ emission
line, the diffuse Ca{\sc ii} bands  and the diffuse interstellar band at 4430
\AA.  The Balmer lines also appear  to be double peaked in minioutburst, with
typical  separations  of  500--700  km  s$^{-1}$,  notably  larger  than  the
separations  observed   during  main  outburst.   The   red  spectra  display
H${\alpha}$ emission and He{\sc i} emission  lines at 5877 and 6680 \AA.  The
He {\sc  i} line at 5877  is affected by the  interstellar blend of  the Na I
D.  The  interstellar  bands  at  $\sim$  5780 \AA  and  6280  \AA  are  also
detected.

The only observations of previous minioutburst spectra of SXTs are those from
GRO  J0422+32 \cite{casares95,callanan95}  and GRS  1009--45 \cite{bailyn95}.
The spectra from  GRO J0422+32 showed broad shallow  H$\beta$ absorption, and
H$\alpha$ evolving from absorption to emission on a timescale of 3 days.  The
spectra from GRS 1009--45 showed weak H$\alpha$ and weak H$\beta$ absorption.
In  none of  them evidence  of He  I  is found.   Casares et  al.  1995  also
obtained phase resolved H$\alpha$ and H$\beta$ spectra of J0422+32 during its
1993 minioutburst.  Balmer lines were  embedded in broad  absorptions whereas
He{\sc ii}  was purely in emission  and showed evidence of  a large amplitude
($\sim$755 km~s$^{-1}$) S--wave component.

\subsubsection{Variability of the emission lines}

\begin{figure} 
\begin{center}
\epsfig{file=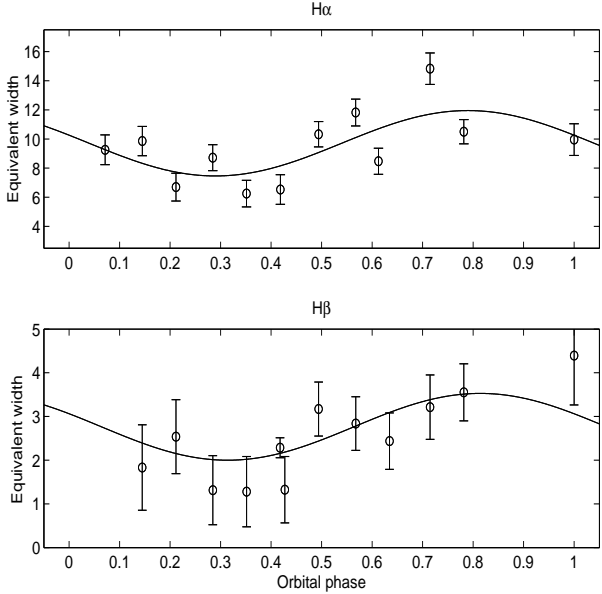,width=8cm,height=8cm,angle=0}
\caption{Equivalent width variations for H$\alpha$ and H$\beta$, during Epoch
B (Jul00 minioutburst).
\label{var_spec}}
\end{center}
\end{figure}

To examine the variability of the emission lines during the minioutburst, the
equivalent widths  were measured in the  individual spectra as  a function of
0.319 d  period (see Figure.~\ref{var_spec}).   The equivalent widths  of the
lines were  measured after normalization  of each individual spectrum  by its
continuum so that  the variations in the lines  are separated from variations
in the  continuum. Unfortunately, this study  was not possible  with the main
outburst  data,  due  to  the  poor  S/N ratio  of  the  individual  spectra.
Sinusoidal fits have  been included in the plots as a  reference.  We can see
that the H$\alpha$ equivalent width  is dependent on orbital phase.  The same
relation is less clear in H$\beta$,  probably due to the smaller amplitude of
the variations,  but it  is still present.   The variation in  the equivalent
widths for both of them can  be roughly reproduced with a sinusoidal function
equal to  the orbital  period of  the system, thus  pointing to  the changing
visibility  of a  region of  enhanced emission  or bright  spot,  perhaps the
splash point  where the gas  stream and accretion disc  meet.  Unfortunately,
the uncertainty in  our T$_0$ prevents to define the  absolute phasing in the
EW curves.  No clear modulation is  seen in the EWs of H$\epsilon$ and He{\sc
ii} which have means of 5.95$\pm$0.95 \AA and 5.41$\pm$1.40 \AA respectively.

%
%

\section{Discussion}

\subsection{Distance estimate}
In the context of King  \& Ritter's model \shortcite{king98}, the exponential
decay of the  X--ray light curve indicates that  irradiation is strong enough
to ionize the  entire accretion disc.  Also, a  secondary maximum is expected
one irradiated-state  viscous time after the  onset of the  outburst and this
can be used to calibrate the  peak X-ray luminosity and hence the distance to
the source $d_{\rm kpc}$ through
$$d_{\rm kpc}= 4.3 \times 10^{-5} t_{s}^{3/2} \eta^{1/2} f^{1/2} F_{p}^{-1/2}
\tau_{d}^{-1/2}$$
\noindent
where $F_p$ is the peak  X-ray flux, $t_s$ the time  of the secondary maximum
after the peak  of the outburst in days,  $\tau_d$ the e-folding time of  the
decay in days, $\eta$ the radiation efficiency parameter and $f$ the ratio of
the disc   mass at the  start of  the outburst to   the maximum possible mass
\cite{sck98}. In our case, $\tau_d$=34 d, $t_{s}\simeq$68 d  and $F_p$ can be
estimated from the XTE count  rate (250 mCrab  in the energy range 2-10  keV)
which corresponds to 1.7  $\times$ 10$^{-8}$ erg cm$^{-2}$ s$^{-1}$. Assuming
$\eta$=0.15 and $f$=0.8 we find $d_{\rm kpc}$ = 11.\\

Alternatively, we  can estimate the distance  to the source  by comparing the
quiescent magnitude with the absolute magnitude of a main sequence star which
fits within the Roche lobe of  a 7.65 hr orbit.  Combining Paczynski's (1971)
expression for the averaged radius of a Roche lobe with Kepler's Third Law we
obtain the  well-known relationship between the secondary's  mean density and
the  orbital  period:  $\bar{\rho}  =  110/ P_{\rm  hr}^{2}$  (g  cm$^{-3}$).
Substituting for the orbital  period of J1859+226 we obtain $\bar{\rho}$=1.87
g  cm$^{-3}$ which  corresponds to  a  K0V-K1V secondary  star with  absolute
magnitude  $M_{R}\simeq$6.1.  The dereddened  quiescent magnitude  is V=23.27
(using  A$_V$=1.80$\pm$0.07 as  derived  from  the NaID  line;  see Hynes  et
al. 1999) which  yields $d_{\rm kpc}$ = 11. This  value perfectly agrees with
that obtained previously,  although strictly speaking, this is  a lower limit
to the distance  as we are neglecting any contribution  by the accretion disc
to  the  quiescent optical  flux.   Although  the  measured (V--R)  color  is
consistent with the calculated density,  a spectral type determination of the
companion star is essential to refine this distance estimate.

\subsection{Relative contribution of the disc}
Following  the color evolution  of the  system, we  can compute  the relative
contribution  of the disc  to the  optical flux  in the  V and  R bands  as a
function of  the assumed spectral type  of the companion  star. Assuming that
the colour  of the secondary  star corresponds to  a normal dwarf of  a fixed
spectral type  and that the  colour of the  disc remains constant  during the
main outburst (as  we can see in Figure~\ref{color}), then  we can relate the
fluxes   of  the   disc  and   star   in  the   R  and   V  bands:   f$_{{\it
V}_{disc}}$=h$\times$f$_{{\it              R}_{disc}}$,              f$_{{\it
V}_{star}}$=k$\times$f$_{{\it    R}_{star}}$,    where   h$\propto$10$^{({\it
V-R})_{disc}}$    and    k$\propto$10$^{({\it    V-R})_{star}}$.     It    is
straightforward to calculate the  relative contribution of the disc (f$_{{\it
V}_{disc}}$/f$_{\it V}$ and f$_{{\it R}_{disc}}$/f$_{{\it R}}$) through:
$$f_{\it V}=h\times f_{\it R}+(k-h)\times f_{{\it R}_{star}}$$
\noindent
where f$_V$  and f$_R$ are  the measured  fluxes in the  V and R  bands.  The
unreddened colours ({\it V-R}) corresponding to spectral types G5V to M0V are
taken from Schmidt--Kaler (1982). We also assume E(B-V)=0.58 from Hynes et
al. 1999. The results are shown in Table~\ref{disc}.

\begin{table}
\caption{Relative contribution of the disc in the R and V band (bold),
assuming different spectral types for the companion star.
\label{disc}}
\begin{tabular}{lcccc}
\hline 
\hline
{\it HJD($^*$)} & {\em G5V}  & {\em K0V}   & {\em K5V}  &  {\em M0V}\\ 
{\it 10}  & {0.99}     & {0.99}     & {0.99}     & {0.99}    \\
      	  & {\bf 1.00} & {\bf 1.00} & {\bf 1.00} & {\bf 1.00}\\
{\it 21}  & {0.99}     & {0.99}     & {0.99}     & {0.99}    \\
          & {\bf 1.00} & {\bf 1.00} & {\bf 1.00} & {\bf 1.00}\\
{\it 65}  & {0.99}     & {0.99}     & {0.99}     & {0.99}    \\
          & {\bf 1.00} & {\bf 1.00} & {\bf 1.00} & {\bf 1.00}\\
{\it 120} & {0.96}     & {0.97}     & {0.98}     & {0.98}    \\
          & {\bf 0.99} & {\bf 1.00} & {\bf 1.00} & {\bf 1.00}\\
{\it 130} & {0.97}     & {0.97}     & {0.98}     & {0.98}    \\
          & {\bf 0.98} & {\bf 0.98} & {\bf 0.99} & {\bf 0.99}\\
{\it 160} & {0.93}     & {0.94}     & {0.95}     & {0.96}    \\
          & {\bf 0.95} & {\bf 0.95} & {\bf 0.97} & {\bf 0.98}\\
{\it 167} & {0.91}     & {0.92}     & {0.94}     & {0.94}    \\
          & {\bf 0.95} & {\bf 0.96} & {\bf 0.98} & {\bf 0.98}\\
{\it 200} & {0.92}     & {0.92}     & {0.94}     & {0.95}    \\
          & {\bf 0.93} & {\bf 0.94} & {\bf 0.96} & {\bf 0.97}\\
{\it 249} & {0.37}     & {0.44}     & {0.55}     & {0.61}    \\
          & {\bf 0.38} & {\bf 0.45} & {\bf 0.56} & {\bf 0.62}\\
{\it 267} & {0.64}     & {0.68}     & {0.73}     & {0.76}    \\
          & {\bf 0.63} & {\bf 0.66} & {\bf 0.75} & {\bf 0.78}\\
{\it 272} & {0.93}     & {0.93}     & {0.94}     & {0.95}    \\
          & {\bf 0.93} & {\bf 0.94} & {\bf 0.95} & {\bf 0.96}\\
{\it 281} & {0.85}     & {0.86}     & {0.89}     & {0.91}    \\
          & {\bf 0.84} & {\bf 0.85} & {\bf 0.88} &{\bf 0.90}\\
\hline 	      
\hline	      
\footnotesize{$^*$HJD--2451460}
\end{tabular} 
\end{table}

During the first  phase of  the   outburst, the  optical  emission is  almost
completely dominated by the accretion disc. The  relative contribution of the
disc reaches a minimum when  the system drops to the  faint level between the
June 00 and July 00 minioutbursts.

\subsection{Properties of the disc}
Using the same technique, we have  fitted the unreddened {\it B}, {\it V} and
{\it R}  band fluxes to  estimate the intrinsic  colours of the disc.   As we
might expect, the fluxes during  outburst are strongly correlated.  We obtain
({\it  B-V})$_{disc}\sim$--0.07  and  ({\it V-R})$_{disc}\sim$--0.10.   These
colours  agree well  with irradiated  model  predictions, where  most of  the
reprocessed energy  is radiated  in the  {\it UV} (see  e.g. van  Paradijs \&
McClintock 1995).\\


The  ratio of  outburst X-ray  to  optical luminosity  [$\xi=B_{0} +2.5  \log
F_{\rm  x} (\mu\rm{Jy})$] agrees  with the  observed distribution  for LMXBs.
Taking B=15.9  (Chaty et  al.  2000)  and F$_{\rm x}  (2-12 keV)  \simeq 250$
mCrab  \cite{wood99}  at the  outburst  peak  and  assuming $A_{\rm  B}$=2.39
\cite{hynes99} we  obtain $\xi$=20.9 $\pm$0.4, whereas  the distribution peak
of LMXBs gives  $\xi$=21.8 $\pm$1 (see van Paradijs  \& McClintock 1995). The
moderately low $\xi$  probably indicates a high binary  inclination since the
X-ray source  in J1859+226 could be  partially hidden by  the accretion disc.
This result  is consistent with the  evidence for two X--ray  dips during the
second outburst ($\sim$July 8) reported by  Tomsick et al. 2000, and with the
large amplitude (0.4 mag) of the quiescent ellipsoidal modulation.

\subsection{The outburst mechanism}

It  is largely  accepted that  optical emission  in SXTs  outburst is  due to
irradiation of the  outer disc by X--rays.  Within  this context, the optical
emission must be correlated with  the X--rays. However, we see features which
have no correspondence  between the X--rays and optical,  such as a secondary
maximum $\sim$12  days after the  peak of the  X--rays outburst which  is not
visible in the optical band, and minioutbursts which are absent in the X--ray
lightcurve. Such differences can not be  explained in the context of a simple
irradiated disc  and an  additional mechanism is  necessary to  explain them.
Systems  with clear discrepancies  between X--rays  and optical  are J1655-40
\cite{esin00}   and  J1550-564  \cite{jain00}.    Both  sources   exhibit  an
exponentially  declining  optical light  curve,  whereas  the X--ray  remains
constant or increases slightly. To explain the outburst light curves in these
systems  it has  been suggested  the classical  dwarf--nova  type instability
followed by  an episode of  enhanced mass transfer  from the secondary  or up
scatter of  the optical flux into  the X--ray by the  corona. Our photometric
colors indicate that the optical emission is dominated by X--ray reprocessing
on the disc although in  J1859+226 clearly interplay with viscous heating and
geometrical effects  (such as disc  height variations and  shadowing effects)
are  playing an  important role  in the  differences between  the  X--ray and
optical lightcurves.\\

The only SXTs where minioutbursts have been detected are XTE J0422+32 and GRS
1009--45, although this property does not  seem to be a peculiarity of these
systems,  but  a  selection  effect,  due  to  the  difficulty  in  obtaining
continuous  monitoring  from  outburst  to quiescence.   The  most  promising
mechanism to produce minioutbursts is the {\it X--ray echo model} (Augusteijn
et al.   1993, Hameury et al.  2000) , where  they are interpreted as  due to
enhanced  mass flow  from  the companion  star  on the  outer  disc which  is
responding,  essentially linearly, to  heating by  X--rays from  the primary.
The process  goes on  continuously, triggered by  echoing the  initial X--ray
outburst.\\

The  spectra taken  during  the main  outburst  show an  emission feature  at
$\lambda\lambda$4540--4550\AA which is  the {\it Bowen Blend}, but  it is not
present in the Jul 00 minioutburst.   This behavior has also been observed in
J0422+32 (Casares et  al. 1995, Callanan et al. 1995).  The  Bowen blend is a
combination of high excitation lines  (mainly C{\sc iii}, O{\sc ii} and N{\sc
iii} at $\lambda\lambda$4634--4642), produced by the fluoresencence resonance
mechanism which initially  requires seed photons of He{\sc  ii} Ly$\alpha$ at
$\lambda$303.78.  Under the hypothesis  that the composition of the accreting
material  has not  changed substantially  since  the onset  of outburst,  the
absence  of Bowen  emission in  the minioutburst  spectra must  be ultimately
related to the weakness of the X-ray photoionizing continuum which originates
the whole cascade process.  We do  not have enough information in our data to
make  detailed calculations  of this  process.   However, it  seems that  the
reprocessed  X--rays make  a small  contribution to  the optical  flux during
minioutburst: in an irradiated  disc, the average ratio of  X--ray to optical
luminosity yields L$_X$(2--11 keV)/L$_{opt}$(300--700 nm)$\simeq$500.  We can
estimate this rate during the highest minioutbursts (Jul 00).  Tomsick et al.
2000 report  a 3.11$\times$ 10$^{-11}$erg  cm$^{-2}$ s$ ^{-1}$ flux  near the
maximum of  the minioutburst in the  range 2.5--20 keV, which  was well above
the expected  quiescent X--ray flux  level.  Assuming the energy  spectrum is
described by a  power law with a  photon index 2.03 as they  reported, we can
estimate the flux in the range 2--11 keV.  Taking a {\it B} magnitude of {\it
B}$\sim$19.2,  the   estimated  X--ray   to  optical  ratio   is  L$_X$(2--11
keV)/L$_{opt}$(300--700  nm)$\sim$4.   This means  that  the X--ray  spectrum
during  minioutburst must  be much  harder than  during primary  outburst or,
alternatively, that the optical flux during minioutburst is instead dominated
by the  intrinsic luminosity  of the disc.  During the primary  outburst, the
intensity and  width of the Balmer  lines is notoriously smaller  than in the
minioutburst spectra as expected because the continuum is 2 mag brighter. The
change   in  line   size  is   probably  reflecting   a  shrinking   in  disc
size. Alternatively  the emission line cores  might be filled  in with narrow
emissions during the main outburst.\\

Another possibility could be  that  minioutbursts are  produced by a  temporal
enhancement of viscosity in the  quiescent disc just  after the main outburst
\cite{osaki97}.   However,  there is  no  easy explanation   as  to  why  the
viscosity can vary in such a way.
%
%
\section{Acknowledgments} 
We thank G. Dalton for contributions to the observing campaign and E. Kuulker
for providing  the optical  light curve  of J0422+32.  Part  of this  work is
based  on observations  made with  the  European Space  Agency OGS  telescope
operated on  the island of Tenerife  by the Instituto  de Astrof\'{\i}sica de
Canarias  in  the  Spanish  Observatorio   del  Teide  of  the  Instituto  de
Astrof\'{\i}sica  de  Canarias.   TS  was  supported by  an  EC  Marie  Curie
Fellowship HP--MF--CT--199900297.\\
%
%

\end{document}